\documentclass[pdflatex,sn-mathphys-num]{sn-jnl}

\theoremstyle{thmstyleone}%
\theoremstyle{thmstyletwo}%

\theoremstyle{thmstylethree}%

\usepackage{adjustbox}
\usepackage{natbib}
\usepackage{amsmath}
\usepackage{float}

\usepackage{xr}
\makeatletter

\newcommand{\overbar}[1]{\mkern 1.5mu\overline{\mkern-1.5mu#1\mkern-1.5mu}\mkern 1.5mu}

\begin{document}

\title[Statistical comparisons of time-series feature sets on classification tasks]{Statistical comparisons of time-series feature sets on classification tasks}


\author*[1]{\fnm{Trent} \sur{Henderson}}\email{then6675@uni.sydney.edu.au}

\author*[1]{\fnm{Ben D.} \sur{Fulcher}}\email{ben.fulcher@sydney.edu.au}

\affil[1]{\orgdiv{Dynamics and Neural Systems Lab}, \orgname{The University of Sydney}, \orgaddress{\city{Camperdown}, \postcode{2006}, \state{NSW}, \country{Australia}}}


\abstract{
In recent years, numerous open-source software libraries have been developed for computing sets of features from univariate time series.
The type and number of features vary across these feature sets, which have been constructed with varying disciplinary perspectives on quantifying structure in time-series data.
To date, the relative strengths and weaknesses of these feature sets on time-series classification problems remains largely unexplored.
Here we aimed to understand the relative performance of six open-source feature sets and three baseline feature sets (based on distributional and/or basic spectral structure) across 124 univariate time-series classification problems using a normalization-based approach to problem-level benchmarking that better indexes the relative strengths and weaknesses of different algorithms compared to prior rank-based approaches.
Despite their dramatic differences in size, composition, and computation time, we found that feature sets performed relatively similarly overall (85.3\% of pairwise comparisons resulted in ties), with the largest feature set, tsfresh, exhibiting the strongest overall performance (29.03\% wins across all pairwise comparisons against other feature sets).
We also highlighted specific problems on which the specific composition of a given feature set gave it a substantial performance advantage or disadvantage, and problems where simple baselines comprised of Fourier coefficients and quantiles were sufficient to achieve strong performance.
Our results demonstrate the need to consider problem-level performance when benchmarking time-series feature sets, and highlight the importance of feature make-up in driving relative classification performance.
}


\maketitle


\section{Introduction}\label{sec:intro}

Time series, repeated measurements taken over time, are an important data type across many disciplines that are used to solve problems such as signal classification and forecasting.
Here we consider univariate time-series data, which can be represented as $\mathbf{x} = (x_{t}, \dots, x_{t+T})$, capturing $T$ uniformly sampled real-valued measurements.
We also focus here on time-series classification---a problem class which aims to map a univariate time series $\mathbf{x}$ to a vector of class labels $\mathbf{y}$ with the goal of predicting the class membership of unseen time-series.
Time-series classification is a common problem in many applied settings---such as astronomy \cite{barbaraClassifyingKeplerLight2022}, medicine \cite{wang2022systematic}, and civil engineering \cite{arul2021applications}.
To ensure that analysts are equipped with appropriate, high-performance methods to solve challenging problems, it is important to evaluate different classification methods against each other (i.e., `benchmarked').

Across different fields of study, hundreds of analysis methods have been developed to understand temporal structure and extract information from time-ordered data.
For example, in econometrics, a researcher may decompose a signal into additive components that can be individually statistically modeled and then combined again using methods such as Seasonal and Trend decomposition using Loess (STL) \cite{cleveland1990stl}; in finance, a researcher may employ generalized autoregressive conditional heteroskedasticity (GARCH) models to predict the movement of a financial instrument whose signal is non-stationary \cite{bollerslev1986generalized}; and in astronomy, a researcher may examine properties of the frequency domain using Fourier transforms \cite{brigham1988fast}.
Can these disparate methods be collectively used to develop a better understanding about a time-series dataset?

One way of leveraging the large number of time-series analysis methods is to extract a real-valued scalar summary statistic from each, known as a `feature'---such as a coefficient of a fitted GARCH model or the maximum of the power spectral density of the signal \cite{fulcherFeatureBasedTimeSeriesAnalysis2018}.
Taking a feature-based approach to time-series analysis reduces the time series $\times$ time matrix to a time series $\times$ feature matrix which can then be used as input to statistical learning algorithms such as classification or regression models \cite{fulcherHighlyComparativeFeaturebased2014, fulcherFeatureBasedTimeSeriesAnalysis2018, tan_time_2021, Guijo_Rubio_2024}.
For time-series classification applications, the problem is to map the $N \times T$ (time series by time) data matrix $\mathbf{X}$ to a vector of class labels $\mathbf{y}$ \cite{faouzi2024time}.

Time-series features derived from underlying scientific theory can provide interpretable insights into the dynamics \cite{fulcherHighlyComparativeTimeseries2013}.
In contrast to machine learning methods whose learning mechanisms are either obfuscated by complexity or unable to be directly interpreted (i.e., they are a `black box', such as neural networks \cite{alain2016understanding}), using time-series features derived from underlying theory provides a level of interpretability that facilitates understanding that can motivate subsequent decision-making and scientific inquiry.
Interpretable methods are also crucial for applications in which transparency is critical; such as medicine or policy making \cite{vellido2020importance, bell2022s, ciobanu2024critical}.

Over the past two decades, numerous software libraries have been developed for extracting sets of time-series features from data.
Each of these sets---while unique in their own composition and application programming interface---comprises a collection of features that aim to capture a range of temporal properties.
These libraries were not designed to implement a specific time-series analysis method but to instead implement a wide variety of methods \cite{theft_r_journal}.
To our knowledge, the most broadly comprehensive existing time-series feature set is \texttt{hctsa} (`highly comparative time-series analysis'), a Matlab library containing $>7700$ features from across the interdisciplinary literature~\cite{fulcherHighlyComparativeTimeseries2013, fulcherHctsaComputationalFramework2017}.
While extensive in its coverage of methods, \texttt{hctsa} requires access to proprietary software (Matlab), which limits its broader scientific adoption.
Since the release of \texttt{hctsa}, six prominent open-source feature sets have been developed which are readily accessible across a range of programming languages:

\begin{itemize}
    \item \texttt{catch22} (C, Matlab, R, Python, Julia)---computes a general subset of 22 features from \texttt{hctsa} which were selected through an optimization procedure which sought to maximize classification accuracy while minimizing feature--feature redundancy \cite{lubbaCatch22CAnonicalTimeseries2019}.
    \texttt{catch22} is often extended to include mean and variance to form \texttt{catch24}.
    \item \texttt{tsfeatures} (R)---computes 62 features from methods commonly used by econometricians and forecasters, such as crossing points, seasonal and trend decomposition using Loess, autoregressive conditional heteroscedasticity (ARCH) models, and unit-root tests \cite{tsfeatures_pkg}.
    Sixteen of its features also come from \texttt{hctsa}, and this collection has been previously employed to organize tens of thousands of time series in the \textit{CompEngine} time-series database \cite{fulcherSelforganizingLivingLibrary2020}.
    \item \texttt{feasts} (R)---computes 43 features from similar domains and statistical methods as \texttt{tsfeatures}, where a subset are shared \cite{feasts_pkg}.
    \item \texttt{tsfresh} (Python)---computes 783 features associated with properties of the autocorrelation function, entropy, quantiles, fast Fourier transforms, and distribution \cite{christDistributedParallelTime2017, christTimeSeriesFeatuRe2018}.
    \item \texttt{TSFEL} (Python)---computes 156 features associated with the distribution, autocorrelation function, spectral quantities, and wavelets \cite{barandasTSFELTimeSeries2020}.
    \item \texttt{Kats} (Python)---computes 40 features (of which 30 are Python implementations of \texttt{tsfeatures} features) associated with crossing points, STL decomposition, sliding windows, autocorrelation and partial autocorrelation function, and Holt--Winters methods for linear trends. \cite{Jiang_KATS_2022}.
\end{itemize}

Our previous work investigated differences between these six feature sets, with a focus on computation speed, within-set feature redundancy, and between-set redundancy and overlap \cite{hendersonEmpiricalEvaluationTimeSeries2021}.
It was found that:
(i) computation time varied over three orders of magnitude between them;
(ii) some sets are quite diverse in their feature composition while others exhibit considerable redundancy; and
(iii) there was substantial overlap between the sets, with \texttt{tsfresh} being the most distinctive feature set (in part due to its incorporation of large numbers of Fourier coefficients across real, imaginary, angle, and absolute components, which are not measured by the other sets or are summarized at higher levels).
While this work was foundational in its evaluation of available time-series feature sets, to our knowledge, no research has formally investigated the relative performance of each feature set on real-world problems.

Benchmarking time-series feature set performance is challenging.
Given the differences in feature composition between existing feature sets, we expect each to have different strengths (where differences between classes of time-series are well-captured by the types of features in a given feature set) and weaknesses (where they are not).
For example, feature sets which comprehensively measure Fourier transform coefficients at high resolution, such as \texttt{tsfresh}, should perform well on problems where the classes differ clearly in specific frequency-domain characteristics (including phase information), whereas feature sets that summarize spectral properties at a coarser level (e.g., as bands of a power spectrum), are `blind' to phase information and would thus be expected to exhibit inferior performance in general.
Strong relative performance for a given feature set can be achieved when it uniquely contains features that can discriminate between classes, but, given the high redundancy between feature sets \cite{hendersonEmpiricalEvaluationTimeSeries2021}, it is unclear whether the specific differences in make-up between feature sets translates into substantial differences in performance.

In addition to differences in feature composition, we also expect characteristics of the data to impact the relative classification performance of feature sets.
For example, a relatively large feature set (like \texttt{tsfresh} with 783 features) can provide a more expansive statistical characterization of time-series structure (than a small set, like \texttt{catch22} with just 22 features) but could be more prone to over-fitting on problems with small numbers of training samples \cite{verleysen2005curse}.
Thus, performance variability of feature sets is driven by multiple factors, including the types of temporal properties each measures as well as the ratio of training data relative to feature set size.

Given these expected strengths and weaknesses, applying aggregated benchmarking methods---which distill performance across datasets to a metric such as an average rank \cite{demsarStatisticalComparisonsClassifiers2006}---is too coarse for the purpose of this work, which is to discern \textit{when} and \textit{why} a given feature set performs well or not.
It is crucial to explore problem-level differences between feature sets and not just aggregate summaries to identify the types of problems where a feature set does or does not contain informative features for distinguishing between classes that the other sets do.
These results can then be interpreted directly by connecting performance to the feature composition of a given set relative to competitors.

Benchmarking problem-level differences between time-series feature sets on a wide range of problems has the potential to inform the way feature sets are used in to solve important data-driven problems.
Previously, it has been relatively common practice to tackle a given data-driven problem by manually selecting one feature set to use, without comparison to alternatives \cite{liu2020sensor, luqian2021human, yang2021anomaly, de2022data, gao2025english}.
However, the subjectivity of selecting only a single feature set raises the question about whether substantially better performance could have been achieved through the usage of a different feature set (whose feature composition may be better suited to the problem at hand).

Here we address two key gaps in the literature: (i) the need for a systematic evaluation of time-series classification performance between feature sets; and (ii) the development of a benchmarking methodology that considers the magnitude (rather than simply rank-level differences) of problem-level differences.
We introduce a statistical benchmarking procedure and use it to investigate the relative performance of six pre-existing feature sets on the UEA/UCR repository of over 120 problems \cite{UEAUCRRepository}, allowing us to characterize their relative strengths and weaknesses across different types of data.


\section{Method}\label{sec:method}

This study aims to benchmark the performance of six existing time-series feature sets and three simple baseline feature sets across a large database of time-series classification problems.
We first describe the datasets within the UEA/UCR repository and our approach to normalizing time-series values prior to the extraction of features in Sec.~\ref{sec:datasets}.
We then describe the procedure for time-series feature extraction (in Sec.~\ref{sec:extraction}), the methodology for computing performance on each dataset (in Sec.~\ref{sec:classificationapproach}), and present our approach to statistical comparisons of performance (in Sec.~\ref{sec:stat-testing}).

\subsection{Datasets}
\label{sec:datasets}

To effectively benchmark time-series feature sets, we required a large database of problems.
The UEA/UCR Time Series Classification Repository \cite{UEAUCRRepository} is such a database, containing a multitude of univariate problems\footnote{Downloaded on 5th May, 2026.}.
Each problem is pre-partitioned into separate train and test sets of varying sizes, with problems such as \texttt{FordB} containing the majority of time series in the train set ($81.8\%$), while others such as \texttt{FreezerSmallTrain} containing the majority in the test set ($99.0\%$).
At the time of download, there were 143 univariate problems in the repository from a range of real-world systems and sources including audio samples, processing engineering sensors, imaging, and electroencephalograms.
Time-series features for eighteen problems were unable to be successfully calculated for all feature sets due to some time series not meeting the minimum length requirement for one or more of the time-series feature sets, and so were removed.
We also removed the problem \texttt{Fungi} due to it only containing one time series per class.
We analyzed the remaining 124 problems throughout the remainder of this paper.

\subsubsection{Dataset normalization}

Normalizing time-series values via a $z$-score transform prior to feature computation standardizes comparisons between feature sets that are sensitive to differences in mean and variance (which are properties that ignore the time ordering of data) and those which $z$-score as a common internal processing step (such as \texttt{catch22}).
In addition, $z$-score transforming time-series values ensures that the predictive signal of classifiers trained on time-series features cannot be driven by trivial and non-temporal statistical properties (i.e., beyond the first two distributional moments) \cite{henderson2023dull}.
This is particularly important for the UEA/UCR repository, which has not consistently normalized time series across problems (and some problems have been normalized incorrectly, such as \texttt{ECG200} where the sum of squares of the series can classify the data perfectly \cite{bagnallGreatTimeSeries2017}).
As such, we $z$-scored all time series on all problems prior to calculating features.

\subsection{Feature extraction}\label{sec:extraction}

Time-series features for all six feature sets were extracted using the \texttt{theft} package for R, Version 0.9.0 \cite{theft_r_journal}.
\texttt{theft} computed features using version 0.2.3 of \texttt{Rcatch22}, version 0.4.2 of \texttt{feasts}, version 1.1.1 of \texttt{tsfeatures}, version 0.2.0 of \texttt{TSFEL}, version 0.21.1 of \texttt{tsfresh}, and version 0.2.0 of \texttt{Kats}.
In addition to the six pre-existing time-series feature sets, we also included three additional simple feature sets to serve as a baseline to determine if the complexity of the features contained within the pre-existing feature sets is required for high performance: (i) a set of 200 fast Fourier transform (FFT) coefficients (the squared magnitude and angle at frequencies $0-99$); (ii) a set of 101 quantiles, equally spaced in $0.01$ intervals between $Q_{0.0}$--$Q_{1.0}$; and (iii) the union of both the quantiles and FFT coefficients sets---the joint set of 301 features including all 200 FFT coefficients and 101 quantiles---as a third, more comprehensive baseline.
The three baseline feature sets were also computed using Version 0.9.0 of \texttt{theft}.

Time series structure can be separated into distribution structure (which ignores time ordering) and structure contained in the temporal ordering of values.
We included simple implementations of both to test the extent to which the complex and diverse feature sets could outperform these fast and highly interpretable baselines.
The FFT baseline set provides an interpretable frequency-space representation of the data, via projection onto a sinusoidal basis \cite{nyquist1928certain, shannon1949communication}.
Our implementation of this baseline set enables some level of comparison to \texttt{tsfresh}, which includes the same 100 FFT angle coefficients, but which instead computes 100 magnitude coefficients rather than the squared magnitudes (i.e., `power' \cite{bingham1967modern}) which we calculated here, in addition to 200 equivalent representations of the same information (in the form of real and imaginary parts)~\cite{christDistributedParallelTime2017, christTimeSeriesFeatuRe2018}.

Quantiles are invariant to the time ordering of the data and instead characterize the shape of the probability distribution \cite{hyndman1996sample}.
Comparing to the performance of the baseline set of quantiles (which captures distributional information) allows us to assess the relative contribution of complex metrics of temporal structure contained in other feature sets.
We additionally include the union baseline set of FFT coefficients and quantiles to evaluate whether the combination of the independent statistical properties captured by each baseline set (distributional and correlation structure) leads to stronger relative performance against the more complex feature sets.

\subsection{Classification approach}\label{sec:classificationapproach}

In this work, we perform time-series classification using ridge regularized multinomial (or binomial in the case of two-class problems) logistic regression \cite{hoerl1970ridge, mcdonald2009ridge} trained on the time series $\times$ feature matrix following feature extraction from a given feature set \cite{fulcherHighlyComparativeFeaturebased2014}.
Prior to fitting the classification model, we first removed any features that contained $>10\%$ missing values.
Missing values were then imputed with zeroes following $z$-score normalization of feature vectors using training set mean and variance (i.e., missing values were imputed at the standardized mean).
Ridge regularization (otherwise known as the `L2 norm') is a useful tool for controlling overfitting and handling multicollinearity because it shrinks weak coefficients towards zero and coefficients of correlated variables towards each other \cite{hastie2020ridge}.
Implementation of ridge regularization was motivated by prior work which found considerable redundancy within the six time-series feature sets \cite{hendersonEmpiricalEvaluationTimeSeries2021}.
We adopted ridge regularization here instead of alternate shrinkage methods such as lasso regularization (L1 norm), because we did not want the classifier to perform feature selection \cite{tibshirani1996regression}.

In this work, for problems in the UEA/UCR repository which contained sufficient samples for each class in the training data, we implemented 10-fold cross-validation to optimize the regularization penalty for each feature set on every resample-problem combination.
Given the differences in redundancy between feature sets, we expected to see smaller penalty values (i.e., stronger regularization since the penalty parameter represents the inverse of regularization strength) for more redundant feature sets such as \texttt{tsfresh} and larger values for feature sets such as \texttt{catch22}.
For problems with insufficient samples for cross-validation, we fixed the penalty parameter to the \texttt{sklearn} default of $C = 1.0$ for simplicity, which represents moderate regularization \cite{scikit-learn, sklearn_api}.
Inspection of the distribution of penalty parameters obtained over all possible cross-validation problems revealed a median penalty of $C_{\text{median}} = 0.36$, however, we note that feature set medians were distributed between the extremes of \texttt{catch22} ($C_{\text{median}} = 2.78$) and \texttt{tsfresh} ($C_{\text{median}} = 0.05$)---a range which the default $C = 1.0$ approximately bisects.

\subsubsection{Quantifying predictive performance} \label{sec:quantifyingperformance}

Numerous metrics exist for quantifying the performance of a classifier, including classification accuracy, balanced classification accuracy \cite{brodersen2010balanced}, area under the curve, and F1 score \cite{vujovic2021classification}.
For consistency and comparability with prior research \cite{bagnallGreatTimeSeries2017, middlehurst2024bake}, here we use simple classification accuracy.
To further aid comparability with prior time-series classification benchmarking work, we also implement the resampling-based methodology and pre-processing steps of Bagnall et al. (2017) and Middlehurst et al. (2024), which consists of:

\begin{enumerate}
    \item Split each problem into $30$ separate train--test splits, each seeded for reproducibility, where the first seed is always the pre-designated train--test split available in the UEA/UCR Repository and where subsequent splits retain the same class proportions as the first
    \item Calculate the mean and standard deviation of each feature in the training set and use them to $z$-score the values for each time series in both the training and test sets to ensure that test data is completely unseen by the classifier
    \item Use the $z$-scored time series $\times$ feature matrix to train a classifier (ridge regularized logistic regression) for each feature set and train--test split combination and calculate classification accuracy.
\end{enumerate}

\subsection{Statistical testing and evaluation}\label{sec:stat-testing}

Statistical comparison of classifiers is central to time-series classification benchmarking as it provides insight into how different the expected performance of two classifiers is under the null hypothesis that there is no difference.
A large volume of work in the field \cite{middlehurst2024bake, ruizGreatMultivariateTime2021, wang2024deep} has adopted the approach of Bagnall et al. (2017) and Middlehurst et al. (2024)---large and systematic time-series classification benchmarking studies which compared a diverse range of classification algorithms.
The benchmarking approach used in these studies can be summarized in three broad steps: (i) aggregate average ranks for each method based on classification accuracy over a database of problems (the univariate UEA/UCR time-series classification repository \cite{UEAUCRRepository}); (ii) place the results in a critical difference diagram---a plot for visualizing the statistical tests of comparative performance between different algorithms--- which organizes the algorithms on a line according to average rank \cite{demsarStatisticalComparisonsClassifiers2006}; and (iii) employ post-hoc pairwise comparisons using one-sided Wilcoxon signed rank tests \cite{wilcoxon1992individual} with the Holm--Bonferroni correction to form `cliques' \cite{garcia2008extension}---groupings on the diagram within which there is no significant difference.

Recent research highlighted a set of limitations with critical difference diagrams, such as non-significant differences between two algorithms in the diagram becoming statistically significant if another algorithm is removed from the analysis \cite{ismailfawaz2023approachmultiplecomparisonbenchmark}.
As discussed in Sec.~\ref{sec:intro}, in the context of this work, there are two additional limitations for benchmarking time-series feature sets: (i) aggregating over problems masks problem-level differences that give insight into why the composition of a feature set may drive below-average or above-average performance on a particular type of problem; and (ii) given the moderate correlations between feature sets \cite{hendersonEmpiricalEvaluationTimeSeries2021}, we expect to see broadly similar performance on average (i.e., each set contains features sensitive to similar underlying properties), and so aggregating to ranks---which are insensitive to magnitude---imposes a false sense of hierarchy on statistically similar results, whilst diminishing the relative value of statistical below-average and above-average performance.
Together, these limitations prescribe the need to adopt a different statistical benchmarking approach which is both sensitive to magnitude and applicable for individual problems.

\subsubsection{The normalized performance score}

It is challenging to interpret relative differences in classification accuracy between time-series feature sets at a problem level across a database of problems which vary greatly in their number of classes, training sample sizes, and time-series lengths.
The UEA/UCR time-series classification problems vary in their number of classes (ranging from two classes to sixty classes) and also their difficulty \cite{bagnallGreatTimeSeries2017, henderson2023dull, middlehurst2024bake}.
Here we address the dilemma of interpreting the differential difficulty of problems through the `normalized performance score' (NPS), which is a $z$-score of performance for a given feature set against that of all feature sets for a given problem.
This problem-based $z$-score normalization facilitates direct comparison between problems, as it normalizes away sources of inter-problem variation in performance (including in difficulty and number of classes).
We define $a_{ij}$ as being the mean classification accuracy for problem $i$ and feature set $j$, $\mu_{i}$ as being the mean classification accuracy across all resamples from all feature sets on problem $i$, and $\sigma_{i}$ as being the standard deviation of classification accuracy values across all resamples from all feature sets on problem $i$.
Therefore, the NPS for time-series problem $i$ and feature set $j$ is calculated as:

\begin{equation}
    \label{eq:normalizedperformscore}
    \text{NPS}_{ij} = \frac{a_{ij} - \mu_{i}}{\sigma_{i}} \,.
\end{equation}

\subsubsection{Pairwise statistical comparisons}

Pairwise statistical comparisons are a useful way to quantify relative performance using the entire distribution of resampled classification accuracy values for each pair of feature sets.
We implemented these comparisons between each pair of feature sets here to complement the findings obtained through the NPS analysis.
Our pairwise comparisons evaluated the null hypothesis that there was no difference in the means of classification accuracy values between two feature sets, using the distribution of values for each set calculated over the set of 30 resamples:
\begin{equation}
    \label{eq:nullhypothesis}
    H_{0} : \bar{\gamma}_{1} = \bar{\gamma}_{2} \,,
\end{equation}
where $\bar{\gamma}_{1}$ is the mean classification accuracy over all resamples for the first feature set of interest, and $\bar{\gamma}_{2}$ is the mean classification accuracy over all resamples for the second feature set of interest.

Central to our pairwise statistical comparisons is the choice of statistic from which $p$-values can be computed.
Traditional statistical tests between the means of two groups for continuous variables---such as the Student's $t$-test---are inappropriate as the resampled nature of the data violates the assumption of independence.
This causes an inflation of Type I errors due to an underestimation of the variance associated with correlations between samples \cite{dietterichApproximateStatisticalTests1998}.
To account for this non-independence, we use the correlated $t$-statistic \cite{nadeauInferenceGeneralizationError2003}:

\begin{equation}
    \label{eq:tstat}
    t = \frac{\frac{1}{n} \sum_{j=1}^{n}d_{j}}{\sqrt{(\frac{1}{n} + \frac{n_{2}}{n_{1}})\sigma^{2}}} \,,
\end{equation}
where $d_{j}$ is the difference in classification accuracy between the two time-series feature sets for the $jth$ resample, $n$ is the number of resamples, $n_{1}$ is the train test size, $n_{2}$ is the test set size, and $\sigma^{2}$ is the variance of all resampled classification accuracy values.
The inclusion of $(\frac{1}{n} + \frac{n_{2}}{n_{1}})$ as a scaling factor for $\sigma^{2}$ acts as an estimator for cross-sample variance.
We then computed a $p$-value for the correlated test statistic using a threshold of $\alpha = 0.05$, where $p < 0.05$ denotes that the feature set with the higher mean performance is the winner for that particular problem and the other the loser, and $p \geq 0.05$ indicates a tie.
We use the \texttt{R} package \texttt{correctR} \cite{correctR} implementation of this statistic.


\section{Results}\label{sec:results}

This section is structured as follows.
In Sec.~\ref{sec:results_mean_perform}, we investigate mean classification accuracy for each feature set across the UEA/UCR repository to build a baseline understanding of performance trends.
Then, in Sec.~\ref{sec:nps}, we analyze the normalized performance score ($\text{NPS}$, Eq.~\eqref{eq:normalizedperformscore}), to understand the relative classification performance of feature sets for each problem and demonstrate the value of benchmarking methods sensitive to the magnitude of performance differences.
Across both Sec.~\ref{sec:results_mean_perform} and Sec.~\ref{sec:nps}, we use `classification accuracy' to describe the mean classification accuracy of a feature set on a given problem over all 30 resamples.
We finally conduct comprehensive pairwise statistical comparisons between feature sets using the full distribution of resampled values in Sec.~\ref{sec:results_pairwise}.
All code to reproduce our results is openly available on GitHub\footnote{https://github.com/hendersontrent/feature-set-classification}.

\subsection{Building a baseline understanding of average feature set performance}\label{sec:results_mean_perform}

\begin{figure}[htb!]
  \centering
  \includegraphics[width = 0.99\textwidth]{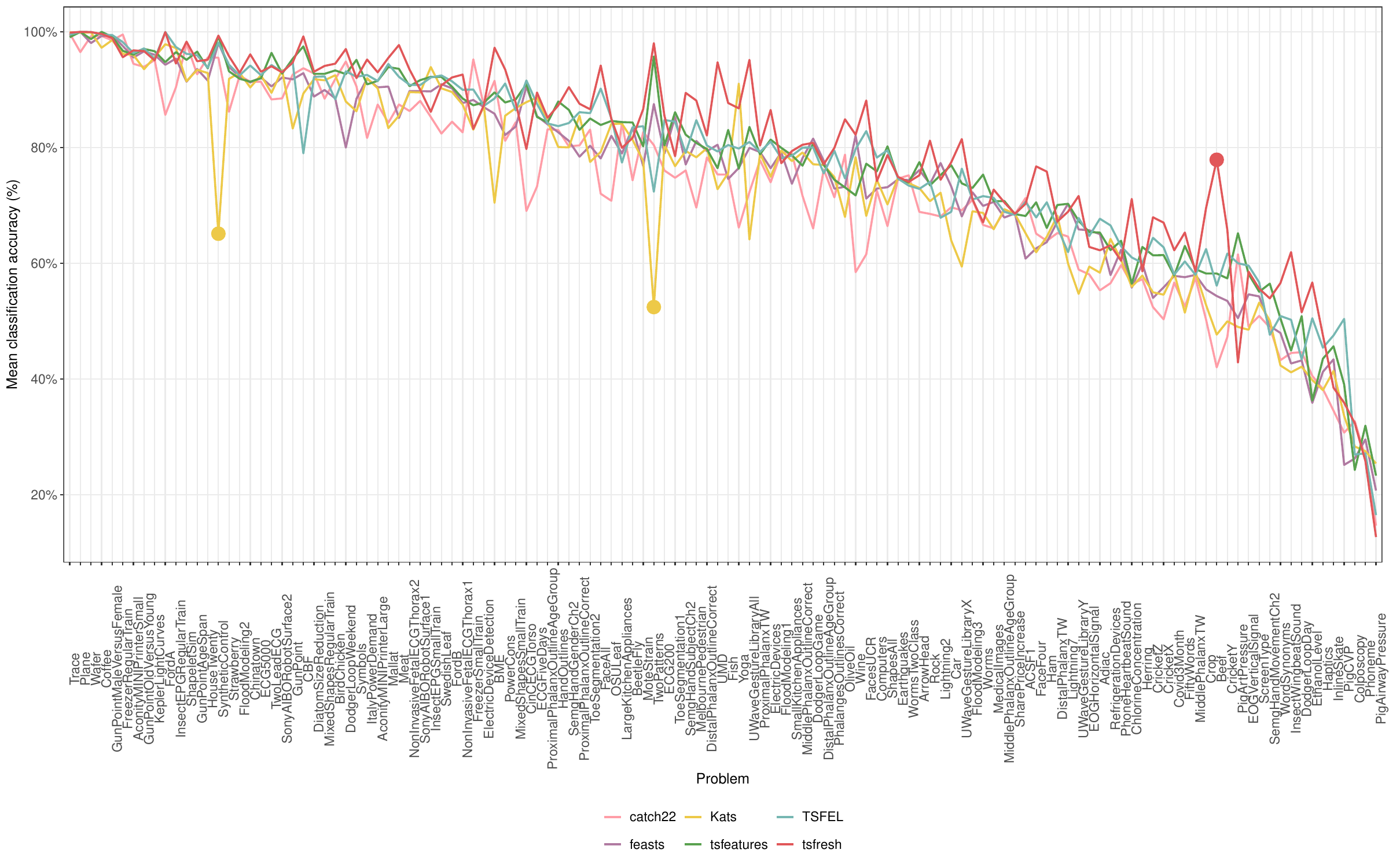}
  \caption{\label{fig:meanplot}
  \textbf{Mean classification accuracy ($\%$) by feature set across problems in the UEA/UCR univariate time-series classification repository.}
Problems are ordered by average classification accuracy across feature sets along the horizontal.
Each of the six feature sets included in the analysis is represented by a colored line.
A selection of three examples in which one feature set exhibits substantial under-performance or over-performance relative to the other feature sets are emphasized by larger circles.
  }
\end{figure}

We first analyzed the classification accuracy of all six time-series feature sets across the 124 time-series classification problems to understand variation in average performance.
Figure~\ref{fig:meanplot} plots the classification accuracy of each feature set for each problem, where problems are ordered by decreasing accuracy across feature sets.
Visually, the different feature sets perform relatively similarly overall, with inter-problem differences in performance clearly dominating variability in performance between feature sets on a given problem, despite large differences in feature comprehensiveness and size between the feature sets.
Specifically, the average variance of raw feature-set accuracy within problems ($\overbar{\text{Var}}_{\text{within}} = 0.003$) was an order of magnitude lower than the average variance of feature set accuracy across problems ($\overbar{\text{Var}}_{\text{between}} = 0.03$).

On top of the broad similarities in feature-set performance, we see a profile of relative strengths and weaknesses for each feature set across problems.
To explore these deviations from the problem average, we highlight three specific problems in which one feature set performed substantially differently to the others (shown as larger points in Fig.~\ref{fig:meanplot}):
(i) \texttt{SyntheticControl} \cite{alcock1999time}, where \texttt{Kats} achieved a substantially lower accuracy of $65.7\%$ (than the average across all other feature sets: $98.1\%$);
(ii) \texttt{TwoPatterns} \cite{geurts2002contributions}, where \texttt{Kats} again achieved a substantially lower accuracy of $55.1\%$ (than the average of all other feature sets: $87.8\%$); and
(iii) \texttt{Beef} \cite{al2002detection}, where \texttt{tsfresh} achieved a substantially higher accuracy of $77.9\%$ (than the average of the other feature sets: $51.7\%$).
On these problems, we aimed to understand the large difference in performance of one feature set relative to the others in terms of its feature composition.

For \texttt{SyntheticControl}, we found that the under-performance of \texttt{Kats} was due to it not containing a feature that is sensitive to linear trend, which is a key statistical difference between the classes in this problem.
\texttt{Kats} instead contains the coefficient of determination value ($R^{2}$) for a linear trend fit to the time series but not the slope value itself \cite{Jiang_KATS_2022}, contributing to its substantial under-performance on this problem.

On \texttt{TwoPatterns}, we found that competitive performance was achieved by \texttt{tsfresh}, \texttt{tsfeatures}, and \texttt{catch22} through time-reversal asymmetry statistics \cite{schreiber2000surrogate, nogare2025identifying}, and by \texttt{TSFEL}, \texttt{feasts}, \texttt{tsfresh} (again), and \texttt{tsfeatures} (again) through measurement of the linear trend.
\texttt{TwoPatterns} is a simulated dataset where four classes differ in the order and direction of stepwise changes (i.e., up--up, up--down, down--up, down--down) in the time-series values.
Time-reversal asymmetry statistics computed on a univariate time series $\mathbf{x} = (x_{t}, \dots, x_{t+T})$ (e.g., $\langle (x_{t+1} - x_{t})^{3} \rangle_{t}$, where $\langle\rangle_t$ represents a mean calculation) from \texttt{catch22} \cite{fulcherHighlyComparativeFeaturebased2014, lubbaCatch22CAnonicalTimeseries2019} capture asymmetries in the distribution of successive decreases in the time series relative to the distribution of successive increases, which makes these features well suited to distinguishing between classes in this problem.
Another type of high-performing feature on this problem---the linear trend coefficient---captures the average directional drift of the time series, meaning that the location of stepwise shifts in the series can drive variation in the sign of the regression coefficient, thus also enabling this feature to distinguish between classes.
As \texttt{Kats} does not contain features sensitive to these statistical properties, it under-performs relative to the other sets.

Finally, we found that the substantial over-performance of \texttt{tsfresh} on \texttt{Beef} \cite{al2002detection}, was driven by its inclusion of a comprehensive Fourier basis function projection (in the form of 400 fast Fourier transform (FFT) coefficients).
For example, when we removed these 400 features from \texttt{tsfresh}, its accuracy dropped by $9.67\%$ ($SD = 5.70\%$), and our baseline feature set containing 200 FFT coefficients (squared magnitudes and phases) also obtained a much higher accuracy on this problem ($72.0$\%) than the other feature sets (average performance of $51.7$\%).
Where other feature sets were constructed to be more compact by summarizing spectral structure at higher levels (such as the variance, skewness, and kurtosis of the frequency distribution computed by \texttt{TSFEL}), a choice that comes at the expense of expressivity, the choice of \texttt{tsfresh} to be a large feature set that retains a comprehensive set of raw Fourier transform coefficients underlies its ability to better capture subtle class differences on the \texttt{Beef} problem.

\subsection{Understanding relative performance using the normalized performance score}\label{sec:nps}

\begin{figure}[H]
  \centering
  \includegraphics[width = 0.945\textwidth]{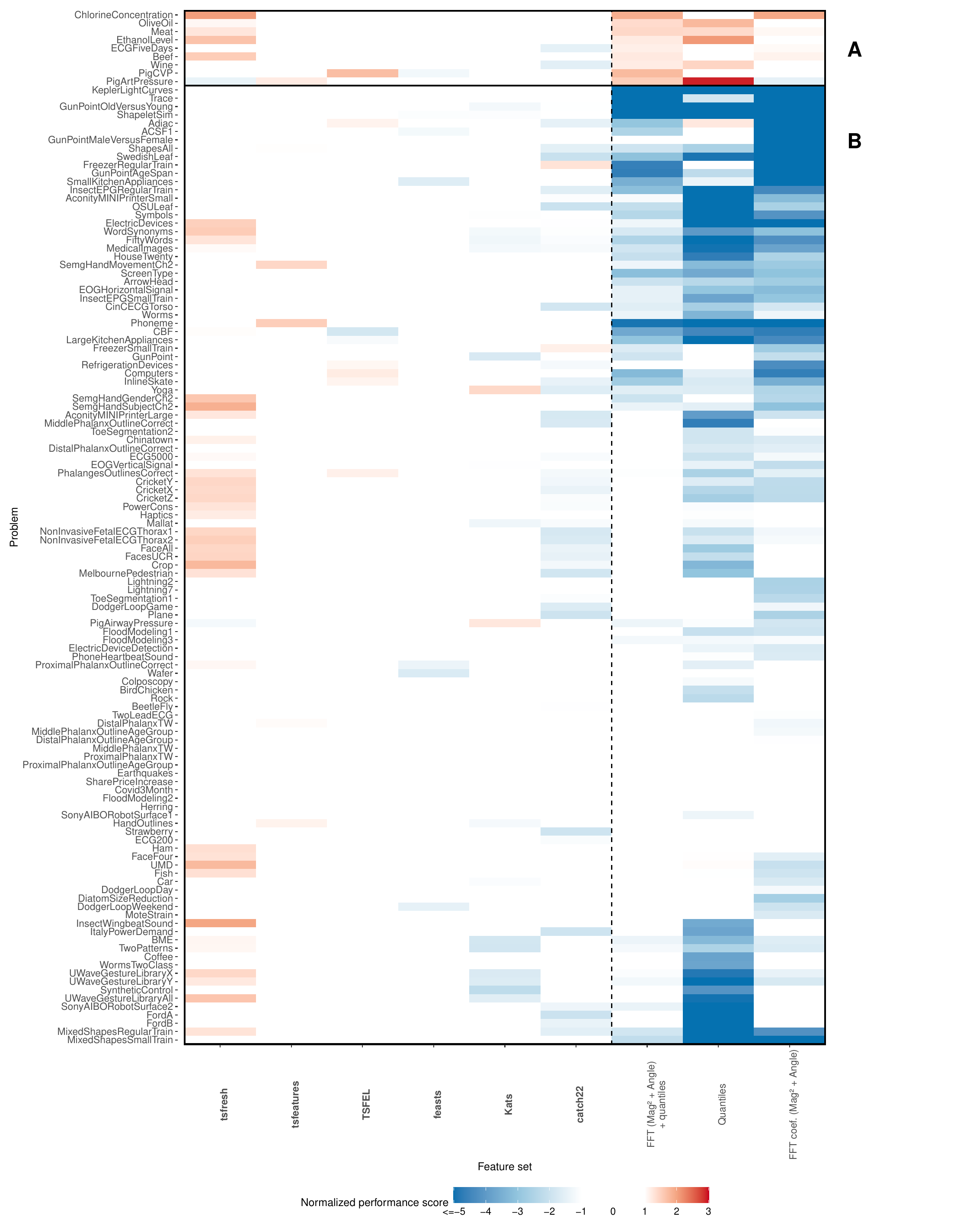}
  \caption{\label{fig:normplot}
  \textbf{Heat map of normalized performance scores (NPS) for each feature set across all problems in the UEA/UCR univariate repository.}
Cell shading indicates the NPS, calculated as a $z$-score across feature sets for a given problem (Eq.~\eqref{eq:normalizedperformscore}).
The feature sets are organized along the horizontal axis in descending order of mean NPS across all problems.
Hierarchical clustering was used to organize problems along the vertical axis for those where baseline feature sets exhibited a normalized performance score less than the mean of the six time-series feature sets.
The NPS is a directly interpretable quantity: $\text{NPS}_{ij} = 1$ indicates that the feature set's mean accuracy was one standard deviation above the mean of all feature sets for the problem.
NPS values in the range $-1 \leq \text{NPS}_{ij} \geq 1$ are shaded white on the heat map to visually emphasize the more distinctive performances.
We annotate to regions of the matrix:
\textbf{A} highlights the collection of 11 problems for which any of the baseline sets achieved a normalized performance score $\text{NPS}_{ij} > 1$, while
\textbf{B} frames the performance of the non-baseline feature sets on the remaining 113 problems where no baseline feature set displayed strong performance, defined as $\text{NPS}_{ij} > 1$.
  }
\end{figure}

To better standardize the relative performance of time-series feature sets across problems which vary in difficulty and compare their relative performance more directly, we analyzed the normalized performance score ($\text{NPS}_{ij}$ of feature set $j$ on problem $i$), computed as a $z$-score across feature set accuracies (excluding the baseline sets) for each problem, where the mean and standard deviation of only the six time-series feature sets was used for the calculation (cf. Eq.~\eqref{eq:normalizedperformscore}).
We present $\text{NPS}_{ij}$ for the set of 124 problems and each of the time-series feature sets and the baseline sets (200 FFT coefficients, 101 quantiles from $Q_{0}$ to $Q_{1}$, and the union of the two introduced in Sec.~\ref{sec:extraction}) as a heat map in Fig.~\ref{fig:normplot}.
Non-baseline feature sets are ordered in the graphic by mean NPS, which we see primarily follows the ordering of feature sets by their size, from the largest (and highest mean NPS) \texttt{tsfresh} through to the smallest (and lowest mean NPS) \texttt{catch22}, with the exception of \texttt{tsfeatures} (62 features) exhibiting a higher mean NPS than the larger \texttt{TSFEL} (156 features).
In order to emphasize cases where there are substantial relative differences in feature-set performances on a given problem, we used white to color any scores within $\pm1$ standard deviation of the problem-level mean (i.e., $\mu_j \pm \sigma_j$ in the notation of Eq.~\eqref{eq:normalizedperformscore}).

Broadly, we see that most of the NPS graphic for the six time-series feature sets ($82.9\%$) is colored white, consistent with the results above (Sec.~\ref{sec:results_mean_perform}) where we found relatively low variability in feature set performance for a given problem.
Despite the large differences in computational complexity between the six time-series feature sets, the fastest feature sets to evaluate (\texttt{catch22} and \texttt{TSFEL}) regularly exhibit similar performance to the more comprehensive and computationally intensive feature sets (such as \texttt{tsfresh})~\cite{hendersonEmpiricalEvaluationTimeSeries2021}.
When the baseline sets are also considered, the percentage of cells colored white decreases to $68.5\%$---a marked reduction compared to only the six time-series feature sets.

We first sought to better understand the problems on which the baseline feature-based representations---which capture the projection onto a periodic basis via the FFT (including measures of correlation structure via the contained squared magnitude coefficients~\cite{wiener1930generalized}), and the shape and characteristics of the probability distribution of time-series values (via quantiles)---achieved a normalized performance score $\text{NPS}_{ij} > 1$.
This allows us to understand problems on which these traditional and generic representations of time-series structure are sufficient to capture the class differences, relative to problems on which more subtle and complex dynamical structures are required.

The set of 101 quantiles outperforms all six dedicated time-series feature sets on four problems: \texttt{PigArtPressure} (with the strongest performance on this problem, $\text{NPS} = 2.92$), \texttt{OliveOil}, \texttt{EthanolLevel}, and \texttt{Meat}.
For these four problems, measuring a comprehensive statistical representation of distributional shape via quantiles, and without any information about the temporal ordering of measurements, is sufficient to achieve strong performance.
The set of 200 FFT coefficients performs $>1SD$ above the mean on four problems (\texttt{ChlorineConcentration}, \texttt{Beef}, \texttt{Meat}, and \texttt{ECGFiveDays}), indicating that class differences are well captured by projecting the data onto a sinusoidal basis.
The union baseline set of FFT coefficients and quantiles performs $>1SD$ above the mean on the same four problems, with the addition of five more driven by the contribution of quantiles: \texttt{PigCVP}, \texttt{PigArtPressure}, \texttt{OliveOil}, \texttt{EthanolLevel}, and \texttt{Wine}.

We now turn our focus to examples where the six time-series feature sets exhibited $\text{NPS}_{ij} < -1$ and $\text{NPS}_{ij} > 1$.
Consistent with its large feature space, \texttt{tsfresh} exhibits the most $>1SD$ performances (39).
We also see that the smallest feature set, \texttt{catch22}, exhibited above average performance on two problems---\texttt{FreezerSmallTrain} ($\text{NPS}_{ij} = 1.17$) and \texttt{FreezerRegularTrain} ($\text{NPS}_{ij} = 1.31$)---as did the 40-feature \texttt{Kats} on \texttt{Yoga} ($\text{NPS}_{ij} = 1.42$) and \texttt{PigAirwayPressure} ($\text{NPS}_{ij} = 1.27$).
These findings demonstrate the ability of the specific composition of these smaller feature sets to achieve strong performance on problems where their unique feature composition can effectively discern between classes, such as the case of \texttt{Kats} measuring `flat spots'---computed as the maximum run length across equal-sized windows along the time series---which accurately distinguishes between classes on the \texttt{Yoga} problem.

\subsection{Pairwise comparisons of feature set performance}
\label{sec:results_pairwise}

We next sought to investigate the rates at which different feature sets statistically outperform each other, assessed using the entire distribution of resampled classification accuracy values.
To achieve this, we implemented our pairwise statistical testing procedure to quantify how often each feature set wins against, ties with, and loses to every other feature set using the corrected resampled $t$-test at a threshold level of $\alpha = 0.05$.
While the potential scope of these comparisons is large, we focus on two key analyses: (i) enumeration of the total number of wins, ties, and losses for every feature set; and (ii) analysis of each pairwise comparison.
We focus on just comparisons between the six time-series feature sets to avoid results being skewed by comparisons to the three baseline sets.

\begin{figure}[H]
  \centering
  \includegraphics[width = 1.0\textwidth]{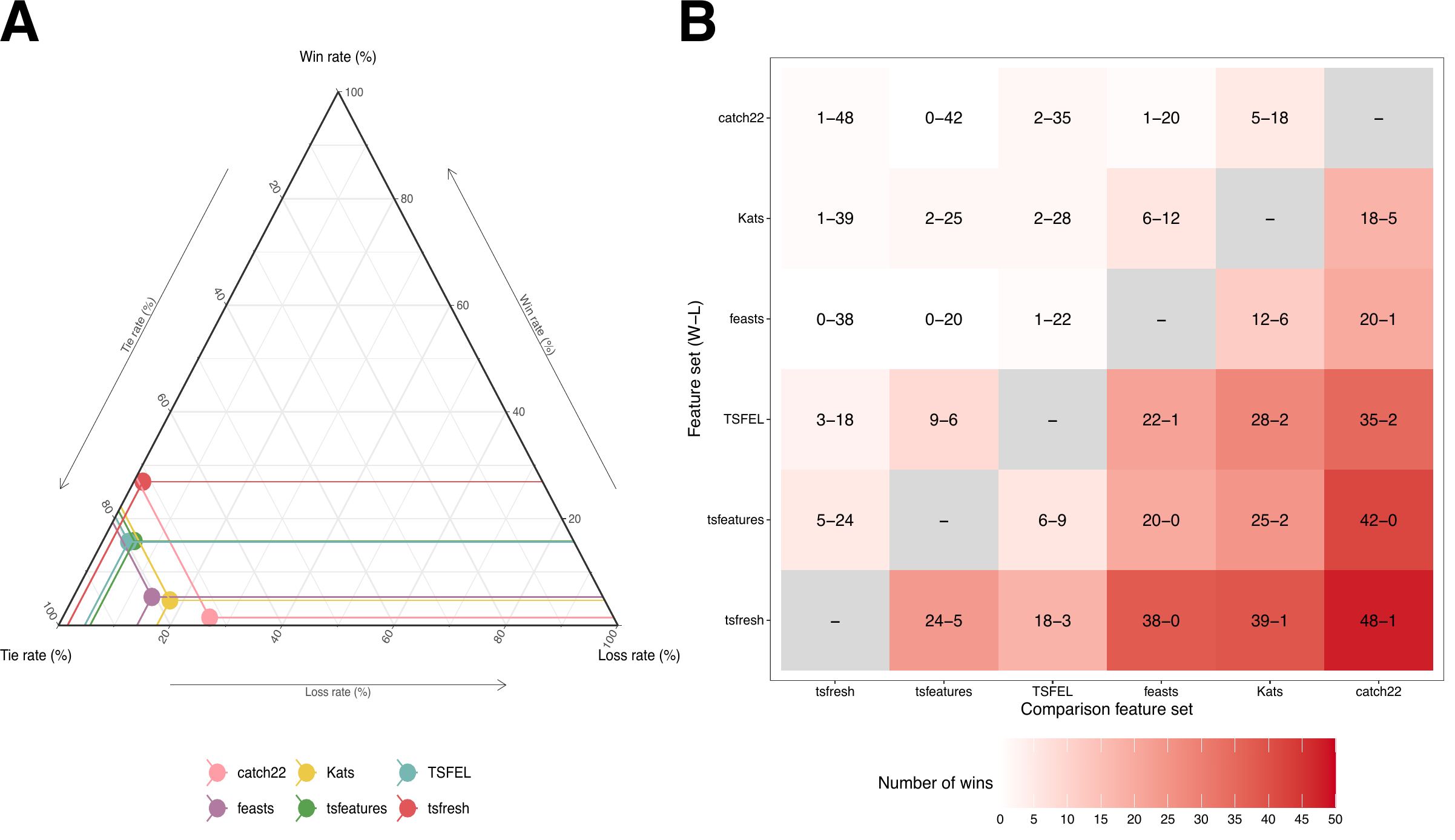}
  \caption{\label{fig:pairwise-summary}
  \textbf{Summary of win, tie, and loss rates for each feature set and head-to-head pairwise comparisons across 124 time-series classification problems.}
\textbf{A} Results from pairwise comparisons are presented in a ternary plot, where tie rate is placed along the left angle axis, win rate is placed along the right angle axis, and loss rate is placed along the horizontal axis.
Each point represents the three-dimensional position of a given feature set according to these metrics.
\textbf{B} Resampled $t$-statistics~\cite{nadeauInferenceGeneralizationError2003} were calculated for each problem for each pairwise combination of the six feature sets.
Number of `statistical wins' (defined as the count of pairwise comparisons with a $p$-value for the test statistic $p < 0.05$ and where the set exhibited the higher mean over resampled classification accuracy values for the problem) is displayed as the first number in each cell and the number of losses as the second.
The remaining cases represent statistical ties ($p > 0.05$).
  }
\end{figure}

We first computed all pairwise comparisons between feature sets across all problems using the method presented in Sec.~\ref{sec:classificationapproach}.
Total win, tie, and loss rates are summarized in Fig.~\ref{fig:pairwise-summary}A, where each outcome is organized along one of the three axes.
We see that feature sets are tightly clustered in the `high tie rate' section of the ternary space: $85.3\%$ of the pairwise comparisons resulted in no statistical difference in mean classification accuracy.
These results reflect the feature--feature correlations which exist between the sets (i.e., there is considerable overlap in the underlying temporal properties each set is sensitive to), which is also consistent with the findings reported in Sec.~\ref{sec:results_mean_perform} and Sec.~\ref{sec:nps} \cite{hendersonEmpiricalEvaluationTimeSeries2021}.

Whilst most comparisons resulted in ties, the largest feature set---\texttt{tsfresh}---wins $29.03\%$ of comparisons, which is far more than the next two feature sets which exhibit nearly identical win rates: \texttt{tsfeatures} ($18.28\%$) and \texttt{TSFEL} ($18.15\%$).
\texttt{tsfeatures} achieves this competitive result with \texttt{TSFEL} with far fewer features, needing only 62 compared to the 156 in \texttt{TSFEL}.

We next investigated the pairwise comparisons underpinning Fig.~\ref{fig:pairwise-summary}A to determine which feature sets perform similarly to which others to gain further insight into the impact of differential feature make-up.
Figure~\ref{fig:pairwise-summary}B displays the matrix of results for all comparisons, where wins--losses are presented as labels in each cell and each is shaded according to number of wins.
In the table, each cell represents the number of times the feature set listed in the row beats the feature set listed in the column.

The highest number of wins was 48 (out of a total 124 problems) for the largest set \texttt{tsfresh} against the smallest set \texttt{catch22}, however, despite vast feature set size and composition differences between these two sets, $60.4\%$ of comparisons between them still resulted in statistical ties.
Win rates against \texttt{tsfresh} were low, as expected, with \texttt{tsfeatures} managing the most (5 wins).
We also see that the smallest win--loss rate was observed for \texttt{tsfeatures} against \texttt{TSFEL}, where only 6 wins for the former and 9 wins for the latter were observed, which is reflective of the overall win--loss--tie rates presented in Figure~\ref{fig:pairwise-summary}A.

There were three cases where a feature set did not beat another: \texttt{feasts} was unable to win against \texttt{tsfresh}, and both \texttt{catch22} and \texttt{feasts} were unable to win against \texttt{tsfeatures}.
The inability of a feature set to achieve a win in a given pairwise comparison suggests that either the comparator set has strong coverage of the benchmark set's features \cite{hendersonEmpiricalEvaluationTimeSeries2021}, or the features unique to the comparator set were better able to distinguish between class differences.
For \texttt{catch22} and \texttt{feasts} against \texttt{tsfeatures}, this finding can be tied to similarity in feature make-up, given that \texttt{tsfeatures} shares a subset of exact features with both \texttt{catch22} and \texttt{feasts}, but is more comprehensive in its coverage through its additional 40 features over \texttt{catch22} and additional 22 features over \texttt{feasts} \cite{theft_r_journal}.

\section{Discussion}
\label{sec:discussion}

In this work, we investigated the relative classification performance of six open-source univariate time-series feature sets \cite{theft_r_journal}: \texttt{catch22} \cite{lubbaCatch22CAnonicalTimeseries2019, Rcatch22_pkg}, \texttt{feasts} \cite{feasts_pkg}, \texttt{tsfeatures} \cite{tsfeatures_pkg}, \texttt{Kats} \cite{Jiang_KATS_2022}, \texttt{TSFEL} \cite{barandasTSFELTimeSeries2020}, and \texttt{tsfresh} \cite{christTimeSeriesFeatuRe2018} using a ridge regularized multinomial (or binomial in the case of two-class problems) logistic regression classifier \cite{hoerl1970ridge, mcdonald2009ridge, hastie2020ridge}.
Our main findings are as follows:
(i) despite differences in the composition of the different feature sets and orders-of-magnitude differences in their size and computational cost~\cite{hendersonEmpiricalEvaluationTimeSeries2021}, feature sets broadly performed similarly to one another on a given problem ($85.3\%$ of pairwise comparisons resulted in no statistical difference), for which more subtle methods for understanding relative feature-set strengths and weaknesses are required to overcome limitations in rank-based aggregations \cite{ismailfawaz2023approach};
(ii) despite the broad trends, on any given problem, feature-set composition can make a substantial difference to performance, with the most numerous set, \texttt{tsfresh}, performing the best on average and recording $29.03\%$ wins across all pairwise comparisons to the other feature sets, and the smallest sets---\texttt{catch22} and \texttt{Kats}---recording performances $>1SD$ above the mean on several problems (such as \texttt{FreezerRegularTrain} for \texttt{catch22} and \texttt{Yoga} for \texttt{Kats}), indicating the usefulness of unique feature composition on specific problems; and
(iii) for several problems, simple baseline feature sets (composed of distributional statistics or Fourier coefficients and the union set of the two) outperformed the dedicated time-series feature sets and achieved competitive performance on many other problems, highlighting the importance of comparison to simple and interpretable methods (that encapsulate the basic properties of distributional shape and correlation structure) to justify the use of more complex approaches \cite{henderson2023dull}.

Despite the broad similarities in classification accuracy, there were notable examples of clear over-performance and under-performance made evident through benchmarking relative performance magnitude.
These cases were able to be tied to the differential make-up of features between the sets, such as the relative under-performance of \texttt{Kats} on the problem \texttt{SyntheticControl} due to it not containing a feature sensitive to trend strength \cite{Jiang_KATS_2022}, which was the most discerning property between the classes.
Our results indicate that while measurement of common statistical properties drives broadly similar performances, feature composition can be a key driver of below-average and above-average performance on any given problem and the uniqueness of individual feature-set combinations can be, in some cases, tied to problem types.

The benchmarking methodology presented in this work deliberately focuses on quantifying the magnitude of relative problem-level differences rather than aggregating performance using average ranks, as has been commonly used in prior algorithmic benchmarking \cite{bagnallGreatTimeSeries2017, thiyagalingam2022scientific, fischer2024large, middlehurst2024bake}.
A common implementation of this aggregation involves computing average ranks via critical difference diagrams with follow-up one-sided Wilcoxon signed-rank tests to determine the top-performing algorithm (on average) across problems \cite{demsarStatisticalComparisonsClassifiers2006}.
More recently, to address some of the limitations associated with critical differences diagrams, supplementary data visualizations---such as accuracy distributions and pairwise plots---have been used to better understand the results in a way that is partially sensitive to magnitude (which rank-based methods are not) \cite{ismailfawaz2023approach, middlehurst2024bake}.
Our results point to key limitations in rank-based comparisons in cases where statistical ties dominate pairwise comparisons, as found here (e.g., $85.3\%$ ties): (i) ranks impose order on performances that are actually not statistically different; and (ii) ranks cannot distinguish cases in which there are significant and substantial differences in performance.
By visualizing relative performance magnitudes via the normalized performance score (Fig.~\ref{fig:normplot}), and using direct pairwise statistical comparison (Fig.~\ref{fig:pairwise-summary}), we were also able to better identify cases of meaningful below-average and above-average performance and then link these cases to the feature composition of the relevant feature sets to understand what statistical properties were missing or uniquely captured to produce the observed results.
Beyond ranking algorithms on aggregated performance, unpicking and interpreting problem-level differences allows us to better understand the relative strengths and weaknesses of different approaches to statistically representing time-series structure on time-series classification problems.
Understanding the specific characteristics of each problem means information that could be used to better tailor methods to data.

Our findings provide a foundation for future feature-based time-series analysis work.
While here we focused on the time-series classification setting, future work may seek to explore comparative performance of the time-series feature sets on other problems like extrinsic regression or forecasting problems \cite{tan2020monash, tan2021time, guijo2024unsupervised}.
In addition, because our research highlighted both the broad usefulness of common statistical properties in driving similar performances, as well as the value of unique features for solving specific problems, a promising future direction would be to construct a new reduced set of high-performing features derived from across the six pre-existing feature sets.
While our results---through the large volume of statistically similar performances---highlight that there may not be substantial gains to changing to an alternate feature set, however, there are cases where performance can be substantially different.
Regardless, deriving bespoke feature sets from a large pool of candidate features (such as the collection of features across all the sets available in the \texttt{theft} package for R \cite{theft_r_journal}) tailored to the input data for a given problem would reduce the subjectivity associated with the selection of a feature set.
The pipeline to create this new feature set could be set up as a multi-objective optimization procedure which balances feature performance (e.g., in a greedy selection procedure \cite{hastie2009elements}), a lack of redundancy (through quantifying feature--feature correlations at each selection stage) \cite{lubbaCatch22CAnonicalTimeseries2019, alam2024canonical}, and computation time \cite{hendersonEmpiricalEvaluationTimeSeries2021}, as well as being able to be tuned to the priorities and needs of a given application.

Lastly, future work may also seek to develop further baseline feature sets beyond our quantiles and FFT coefficients sets---many scientific fields may find value in having a set of tailored baselines specific to the data types commonly analyzed in the field.
Our results show that simple baselines (in this case, a basic, but comprehensive statistical representation of distributional shape and and Fourier coefficients) can outperform more complex and diverse feature sets on some problems, while being clearly interpretable and fast to compute.
Through adoption of appropriate baselines, algorithmic research in the time-series literature will be better able to determine when a new or more complex method adds sufficient performance benefit to justify the tradeoff in interpretability or computation speed.


\section*{Declarations}

\subsection{Competing Interests}

The authors declare no competing interests that may directly or indirectly impact this work.



\bibliography{sn-bibliography}

\end{document}